# Network analysis reveals a potentially 'evil' alliance of opportunistic pathogens inhibited by a cooperative network in human milk bacterial communities

Zhanshan (Sam) Ma[*,1]   Qiong Guan[1]   Chengxi Ye[1,2]
Chengchen Zhang[3]   James A. Foster[4]   Larry J. Forney[4]

[1]Computational Biology and Medical Ecology Lab
State Key Laboratory of Genetic Resources & Evolution
Kunming Institute of Zoology
Chinese Academy of Sciences
Kunming, 650223, China
[*]Correspondence: samma@uidaho.edu

[2]Department of Computer Science
University of Maryland
College Park, MD 20740, USA

[3]School of Public Health
Columbia University
722 West 168th Street
New York, NY 10032, USA

[4]Institute of Bioinformatics and Evolutionary Studies &
Department of Biological Sciences
University of Idaho
Moscow, ID 83843, USA

## Abstract

The critical importance of human milk to infants and even human civilization has been well established. Although the human milk microbiome has received increasing attention with the expansion of research on the human microbiome, our understanding of the milk microbiome has been limited to cataloguing OTUs and computation of community diversity indexes. To the best of our knowledge, there has been no report on the bacterial interactions within the human milk microbiome. To bridge this gap, we reconstructed a milk bacterial community network with the data from Hunt et al (2011)[1], which is the largest 16S-rRNA sequence data set of human milk microbiome available to date. Our analysis revealed that the milk microbiome network consists of two disconnected sub-networks. One sub-network is a fully connected *complete graph* consisting of seven genera as nodes and all of its pair-wise interactions among the bacteria are facilitative or cooperative. In contrast, the interactions in the other sub-network of 8 nodes are mixed but dominantly cooperative. Somewhat surprisingly, the only 'non-cooperative' nodes in the second sub-network are mutually cooperative *Staphylococcus* and *Corynebacterium*, genera that include some opportunistic pathogens. This potentially 'evil' alliance between *Staphylococcus* and *Corynebacterium* could be inhibited by the remaining nodes who cooperate with one another in the second sub-network. We postulate that the 'confrontation' between the 'evil' alliance and 'benign' alliance in human milk microbiome should have important health implications to lactating women and their infants and shifting the balance between the two alliances may be responsible for dysbiosis of the milk microbiome that permits mastitis.

**Keywords**: Human milk microbiome; Network analysis; *Staphylococcus*; *Corynebacterium*; Cooperation, Competition

# Introduction

Human milk is generally considered the best source of nutrients for infants, and its health benefits such as prebiotics, immune proteins, and the microbiome of human milk itself, have been increasingly recognized[1-5]. Similar to other habitats in or on the human body such as the gut and skin, human milk is not sterile at all and it hosts extensive bacterial communities that are posited to possess important health implications. In general, traditional literature on human milk has been focused on pathogenic bacteria, and our understanding on commensal bacteria is still very limited in spite of the rapid advances in metagenomic technology and expanding studies of the human microbiome in recent years. For example, Heikkila & Saris (2003) investigated potential inhibition of *Staphylococcus aureus* by the commensal bacteria of breast milk[6]. *Staphylococcus aureus* is known as a food-poisoning agent and a common cause of infections including serious antibiotic-resistant hospital infections[6,7]. In addition it has been implicated in SIDS (Sudden Infant Death Syndrome)[8,9] as well as infectious mastitis that affects 20-30% lactating women[6,10,11].

There have been several studies that applied metagenomic sequencing technology to characterize human milk bacterial communities[12-17] and a recent one by Hunt et al (2011) provides the largest data set of 16S rRNA sequences from human milk samples[1]. Hunt et al (2011) collected 47 samples from 16 breastfeeding women (3 samples from all but one individual) who self-reported as healthy and between 20-40 yr of age[1]. Their study revealed that the most abundant genera in the milk samples were *Staphylococcus, Streptococcus, Serratia, and Corynebacteria*, while eight other genera had relative abundances exceeding 1%. Besides characterizing the composition of milk bacterial community, Hunt et al. (2011) for the first time described the within-individual variation and the among-individuals variation of milk bacterial communities. The within-individual variation, which can be thought as a measure of the stability of individual milk bacterial communities, differed between individuals. In other words, temporal variation of community membership or stability of individual milk communities varied significantly between women. For example, from the samples of "Subject #5," *Staphylococcus* occupied either the first or second position in terms of the relative abundance (22-59%); but the samples of "Subject #1", *Staphylococcus* only contributed less than 5% to the community abundance. The among-individual variations in the relative abundances of bacteria were as large as six-fold[1].

Although milk microbiome was apparently missing in the initial US-NIH HMP roadmap, its critical importance to human health and diseases is evident. The importance is even more obvious from the perspective of its relationships with the microbiome in other body sites because human microbiome is location specific, but not isolated from one another at all[5,18]. Zaura et al. (2014) hypothesized that development of fetal tolerance toward the microbiome of the mother during pregnancy is a major factor in the successful acquisition of a normal microbiome[19]. Jeurink et al. (2013) proposed a mechanism for the formation of breast milk microbiome, which involves immune cell education by the pregnancy hormone progesterone leading to the transportation of bacteria from the mother to her mammal glands[20]. Guts of breastfed infants showed significantly higher counts of bifidobacteria and *Lactobacillus* and lower counts of *Bacteroides, Clostridium* coccoides group, *Staphylococcus*, and Enterobacteriaceae, as compared with formula-fed infants[21]. The pioneering colonizers such as *Bifidobacterium longum*, which carries several gene clusters dedicated to the metabolism of human milk oligosaccharides (HMOs) allows infants to digest breast milk and possibly some simple vegetal food such as rice. González et al. (2013) found that women with HIV RNA in breast milk have a different pattern of microbial composition, compared with milk without HIV RNA, indicating specific immunological phenomena in HIV-infected women[22]. They also argued that breast milk and infant gut microbiota are essential for the maturation and protection of infant's immune system. A metagenomic study conducted by Ward et al (2013) confirmed the benefits of breast milk ingestion to the microbial colonization of the infant gut and immunity[4]. The latter is demonstrated by the existence of immune-modulatory motifs in the metagenome of breast milk[4]. These recent studies exhibited the significant importance of breast milk microbiome in health and diseases.

Obviously, Hunt et al (2011)[1] and other previous culture-independent studies[12-17,22-23] have deepened our understanding of bacterial communities in human breast milk. In a recent study, Guan and Ma (2014) applied Taylor's power law and neutral theory to investigate the abundance distribution pattern and the maintenance mechanism of milk microbial community diversity, respectively by reanalyzing the existing data on milk microbiome[24]. The analysis with Taylor's power law model indicated that bacterial population abundance in human milk microbiome is aggregated, rather than random, and it was found that neutral theory did not fit to any of the 47 samples (communities), suggesting non-random interactions in community assembly and diversity maintenance[24]. Nevertheless, due to the limitation of the analytical approaches used in previous studies, we still have little knowledge on the bacterial interactions, beyond their non-random nature, within milk microbiome. Indeed, most statistical approaches are not powerful enough to reveal the interspecies interactions within a microbial community[25, 26]. In this article, we take advantage the power of network analysis in studying interspecies or inter-OTU (Operational Taxonomic Unit) interactions in a complex network setting such as a microbial community.

Erdos and Renyi (1960)[27] seminal research on *random graph theory* opened one of the most exciting new fields in combinatorial mathematics in the 20th century, and their random graph theory attracted extensive studies during the subsequent decades. Still, the avalanche of approaches to network analysis and the emergence of network science, of which random graph theory forms a theoretic foundation, was not triggered until the publication of two independent seminal papers published by Watts & Strogatz (1998) on the dynamics of *"small-world"* networks[28] in the journal *Nature*, and by Barabási and Albert (1999) on the emergence of scaling (*i.e.*, *scale free networks*) in random networks[29] in the journal *Science*, respectively. A commonality of both the papers is the extension of basic random graph models so that they can better fit the patterns exhibited by many empirically observed networks in social, technological and natural networks. One of the most active application fields of network science is biology, thanks to the vast datasets available from genomic and metagenomic research. The case for applying network analysis to investigate the human microbiome was argued convincingly by Foster et al. (2008)[30] and since then several network analyses have been successfully performed with human microbiome data[31-33]. One huge advantage of network analysis is its power to visualize multivariate relationships generated from big data sets such as genomic and metagenomic sequence data. Furthermore, various parameters computed with network analysis software packages (*e.g.*, Cytoscape[34], Gephi[35]) offer informative insights on the patterns in biological data. Complex network alignment algorithms and software (*e.g.,* Graphcrunch2[36]) can further be utilized to compare biological networks under different treatments.

## Materials and Methods

The 16S rRNA sequence data sets of human milk were collected by Hunt et al (2011)[1]. Specifically, the V1-V2 region of the bacterial 16S rRNA gene was amplified from genomic DNA using universal primers and approximately 300,000 reads were generated from the barcoded pyrosequencing of amplicons from 47 samples. After quality control, the data set was reduced to approximately 160,000 reads, with a mean of 3400 sequences per sample. The sequence data were assigned to the most likely bacterial genera using the RDP Bayesian classifier. A table of the 15 most abundant genera in each sample was supplied in the Supporting Information (Table S1) of Hunt et al (2011)[1] and was used for our network analysis.

## Results and Discussion

**Results**

The pair-wise relationships among 15 genera were measured by *Spearman rank correlation* coefficients with *p*-value of 0.05, and the computed values of Spearman correlation coefficients with *R-statistics* package (www.r-project.org) were feed into *Cytoscape* network analysis software[34] and *Gephi*[35]. The *Graphcrunch2*[36] software was applied to further compare the reconstructed milk microbiome network with several standard models of complex networks. The results are exhibited in the following Table 1 and Figure 1.

**Table 1. Topological properties of human milk bacterial network**

| Number of nodes | Number of edges | Avg. number of neighbors | Clustering coefficient | Connected components | Network diameter |
|---|---|---|---|---|---|
| 15 | 45 | 6 | 0.944 | 2 | 2 |
| Average path length | Network density | Modularity | Number of communities | Small-world networks | Scale-free network |
| 1.082 | 0.429 | 0.498 | 2 | Yes | No |

`

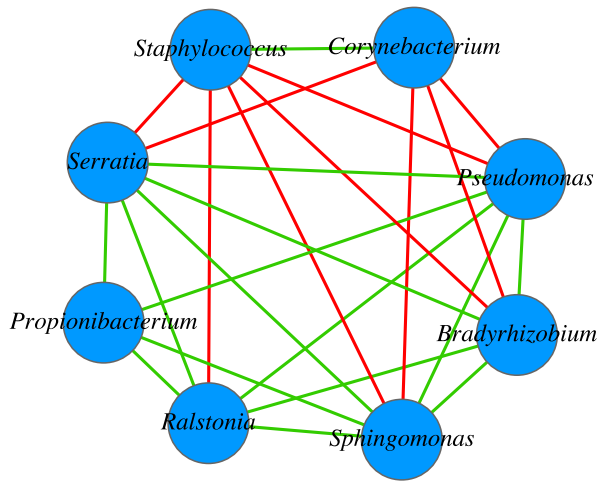
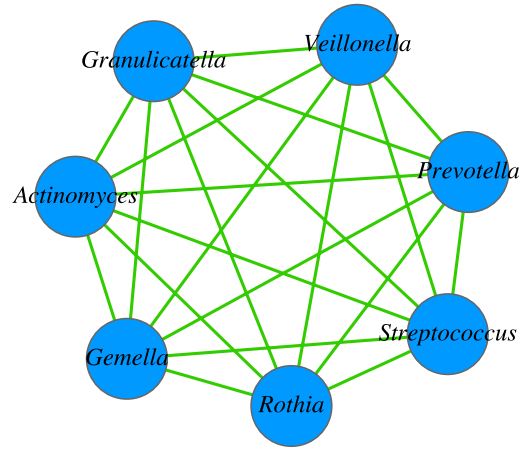

Figure (1a)                                   Figure (1b)

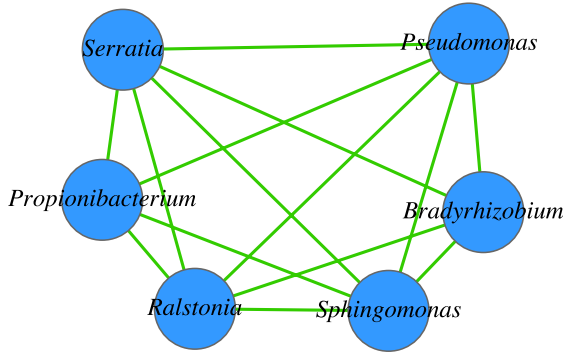
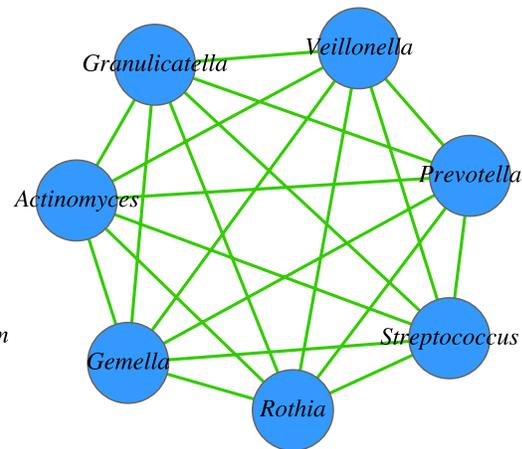

Figure (1c)                                   Figure (1d)

**Figure 1**. Bacterial network of the human milk microbiome reconstructed using the data sets of Hunt et al. (2011): Figure (1*a*) and (1*b*) show the two disconnected sub-networks (components) of the breast milk bacterial network; Figure (1*c*) and (1*d*) are the same components, corresponding to (1a) and (1b), respectively, assuming the *Staphylococcus* and *Corynebacterium* nodes were eliminated. The green line represents a positive correlation (cooperative interaction) while the red line represents a negative correlation (non-cooperative interaction). Obviously, when the two mutually cooperative players *Staphylococcus* and *Corynebacterium* are removed, the whole network becomes totally cooperative (Figure 1c & 1d).

Before discussing our results further, it should be noted that like many other studies of biological networks, our network was built based on the correlation between OTUs. Our usage of terminology such as 'cooperative', 'non-cooperative', and 'evil alliance' are by analogy. Correlation is not equivalent with causation. For example, the correlation between two OTUs may be due to indirect 'facilitation' or 'inhibition' by a third player. Nevertheless, at this stage, correlation data are the only available data type for the human milk microbiome. The correlation network therefore offers the best apparatus available today to tackle the tangled microbiota in human milk. It should also be noted that the OTU data we used resolves taxa at the level of genera and a network based on species data may reveal different patterns from the networks we obtained. Hence, the total validity of the conclusions we draw below, ultimately, should be subject to testing in future biomedical research. Nevertheless, given the fundamental importance of studying milk microbiome, we believe that preliminary analyses such as this are warranted.

From the above results in Figure 1 and Table 1, we draw the following conclusions:
(*i*) The network of the breast milk bacterial community consists of two sub-networks, that correspond to two disconnected components or communities (Figure 1a, Figure 1b, and Table 1). One component (Figure 1b) consists of seven nodes in which all of the nodes are fully connected (forming a complete graph) and they interact cooperatively (positive correlations). Another component consists of 8 nodes and all of the interactions (edges) are cooperative except those that involve *Staphylococcus* and *Corynebacterium*. Therefore, if both *Staphylococcus* and *Corynebacterium* were removed, then the remaining nodes in the sub-network are fully cooperative. Furthermore, the relationship between *Staphylococcus* and *Corynebacterium* are cooperative, although they do not cooperate with other taxa in the network. Obviously, the milk bacterial network is dominantly cooperative, and the ratio of cooperative *vs*. non-cooperative interactions is 4:1.

As mentioned previously, *Staphylococcus aureus* is a food-poisoning agent and a common cause of infections including serious antibiotic-resistant hospital infections[6,7], and the bacterium is also implicated in SIDS (Sudden Infant Death Syndrome)[8,9] and infectious mastitis[6,10,11]. Mastitis is inflammation of the breast with or without infection, and *Staphylococcus aureus* has traditionally been believed to be the pathogen that is typically associated with infectious mastitis[5]. There are studies that reveal some other species of *Staphylococcus* such as *Staphylococcus epidermidis* may play a prevalent role in mastitis infections[2,5,37,38]. Although the etiology of mastitis may vary there is evidence from animal studies and clinical trials that suggests certain strains of *Lactobacillus* can produce anti-inflammatory and anti-bacterial factors that inhibit adhesion and internalization of *Staphylococcus* spp. For example, Arroyo et al. (2010) reported that treatment with probiotic strains from human milk (containing *Lactobacillus* strains) produced a greater reduction in the bacterial counts of *Staphylococcus epidermidis, Staphylococcus aureus* and *Staphylococcus mitis,* as well as a greater reduction in pain than the treatment with antibiotics[39].

The genus *Corynebacterium* includes Gram-positive, rod-shaped bacteria that are largely innocuous and widely distributed in nature. However, some species such as *C. diphtheriae* may cause human disease. Extensive studies on the health implications of breast milk include protecting infants from diarrheal and respiratory diseases[1,40,41]. Our finding from network analysis that every other 'player' in the milk microbiome collectively 'opposes' or 'inhibits' *Staphylococcus* and *Corynebacterium* possibly explains the health effects of milk despite the presence of potential pathogens. In other words, from the perspective of a lactating mother and baby, it may be the cooperative and collective efforts of the other community members that suppress the 'evil' alliance from opportunistically causing unhealthy consequences such as mastitis.

(*ii*) The milk bacterial network is a 'small-world' network with a network diameter of only 2 and average minimal path length of *p*=1.082 (Table 1). This is in accord with the criterion for judging small world networks[28], in which *p*=1.082<log(*N*)=log(15)=2.708. The criterion means that the typical distance *p* between two randomly chosen nodes grows proportionally to the logarithm of *N*, which is even smaller than the linear growth. In biological terms the interactions amongst the nodes (genera) are very tight.

Nevertheless, the milk microbiome network is not *scale-free*, which is evidenced by the fact that the degree distribution of the network does not fit to a power law distribution (*p*-value=0.323>0.05 from fitting power-law model, *i.e.*, $p(k) \propto k^{-\lambda}$). A scale-free network has two properties: growth and preferential attachment[29]. Growth means that the number of nodes in the network increases over time and preferential attachment means that the more connected a node is, the more likely it is to receive new links. This absence of scale-free property may be due to the fact that our network is built with bacterial genera as the node units, which might be relatively constant over time and therefore the growth of nodes could be insignificant. We also utilized the network alignment software *GraphCrunch2*[36] to compare our milk microbiome network with 50 random instances of each of the following network models with the same size as microbiome network, respectively: ER (Erdos-Rényi random graphs), ER_DD (Erdos-Rényi random graphs with the same degree distribution as the data), GEO (Geometric Random Graphs), GEO-GD (Geometric Gene Duplication Models), SF (Scale-free Barabási-Albert Preferential Attachment Models), SF-GD (Scale-free Gene Duplication Models), and STICKY (Stickiness-index Based Models). The results from *graphcrunch2* software also demonstrated that a scale-free network model was among the worst performing models.

(*iii*) The other topological parameters of milk bacterial network shown in Table (1) also offer some interesting information about the network characteristics. Both the numbers of *connected components* and the number of *communities* in the network are two, corresponding to the two sub-networks (Figure 1a & 1b). In general, the values of these two parameters are not necessarily equal, but their equality in the case of milk microbiome further strengthens the evidence that the milk microbiome network is divided into two separate sub-networks. The high clustering coefficient of 0.944 signals the high aggregation tendency of network nodes. Moreover, a network density of 0.429 suggests that the number of edges in the network is about 43%, *i.e.*, less than half of all possible edges with a completely connected network (complete graph). Obviously, the lower density is because there is no edge (connection) between the two disconnected components, which significantly lowers the network density.

**Discussion**

In the study done by Hunt et al (2011)[1] universal primers were used for amplification of the 16S rRNA gene, which increased sequencing coverage and helped them to obtain the most comprehensive experimental survey of milk microbiome to date. Their study confirmed that *Staphylococcus* and *Corynebacterium*, identified here appear to exist as an "evil alliance" in the milk microbiome based on our network analysis even though they are typically present on adult skin too[28]. Hunt et al (2011) had recognized the possibility of contamination from the skin microbiome and took extraordinary caution during sample collection. Furthermore, they compared the compositions of both milk and skin communities and concluded that bacterial communities in milk cannot simply be a result from skin contamination[1]. Some species of *Staphylococcus* and *Corynebacteria* are opportunistic infectious agents, and *Staphylococcus aureus* is associated with lactational mastitis. Other studies have also demonstrated the occurrence of *Staphylococcus aureus* in human milk[1,22,42,43]. It has been reported that, during the course of lactation, up to 30% of women suffer from breast infections or inflammation (mastitis) that often lead to fever, redness, swelling and breast pains[44, 45]. In Hunt et al (2011) study, the milk donors were self-reported as healthy, but at least one of the subjects showed symptoms of mastitis[1]. Therefore, at least, existing literature on milk microbiome referenced above indicates the existence of potentially opportunistic pathogens in milk of both healthy and diseased (mastitis) women. But why do those potential pathogens often seem harmless to lactating mothers and infants? In contrary, the existing literature documented the benefits of milk microbiome such as the protective effects of breastfeeding against diarrheal and respiratory disease as well as reduced risk of developing obesity in infants[1]. There is not an existing theory with sufficient evidence to explain the natural phenomenon in the existing literature of human milk microbiome.

A recent study by Urbaniak et al. (2014) investigated the existence of microbiome within mammary tissue by using 16S rRNA sequencing and culture[46]. They analyzed the breast tissue from 81 women with and without cancer in Canada and Ireland, and confirmed the existence of both health-conferring bacteria such as *Lactobacillus* and *Bifidobacterium*, as well as taxa known for pathogenesis such as *Enterobacteriaceae*, *Pseudomonas*, and *Streptococcus agalactiae*. Yet, none of the 81 women recruited had any clinical signs or symptoms of breast infection. This echoes the phenomenon of the presence of opportunistic pathogens in breast milk revealed by our network analysis in this article.

Our network analysis revealed the possible existence of an "evil alliance" between *Staphylococcus* and *Corynebacterium*, and this alliance is collectively 'opposed' or inhibited by the other members of breast milk bacterial community (network). Our finding from network analysis therefore offers a piece of concrete evidence to support the following hypothesis: Similar to natural ecosystems, the ecosystem of human milk microbiome, which consists of the milk microbial community and its environment (*i.e.*, the human body or host), contains microbial species of various characteristics or functions, being beneficial, harmful, or neutral from a human health perspective. This dynamic balance in the milk microbiome ecosystem depends on the species interactions within the microbiome bacterial community as well as the host environment such as immune system. The states of the milk microbiome (network), which could correspond to healthy or disease states of human body, depend on the interactions within the microbiome network (as exhibited by Figure 1) as well as the host, which may have her unique genomic, immunological, physiological and demographic properties. Specifically, we postulate that in healthy state, the adverse consequences of the "evil alliance" of *Staphylococcus* and/or *Corynebacteria* is 'contained' or inhibited by the collective cooperative defense of other community members. In the face of a disturbance of 'evil alliance' cannot be contained, and the balance may be disrupted or shifted to an alternative state, which may correspond to disease such as mastitis, whose underlying causes or triggers are not known yet. In other words, the "evil alliance" of *Staphylococcus* and *Corynebacterium* are opportunistic and they cause disease to their host only if the collective containment by other members of the microbiome coupled with the defense of host immune system are weakened to such extent that their disturbance cannot be 'absorbed' or tolerated. In a recent study, Ward et al. (2013) postulated that it might be the diversity or genetic traits of the milk metagenome that confer benefits to infants, rather than the action of any one bacterial genus or species[4]. We argue that our hypothesis of dynamic balance of milk microbiome ecosystem is consistent with the ideas postulated by Ward et al. (2013)[4] for the following two reasons: (*i*) the diversity and/or sequences of DNA within the metagenome referred by Ward et al. (2103)[4] are carried by the microbiome, and they are the 'surrogates' of the species or OTUs within microbiome; (*ii*) ecological theory tells us, that diversity and stability, are tangled together, even if stability is not determined by diversity. Indeed, the diversity-stability paradigm debates have been going on for more than a half-century now, but nobody denies their close relationship. In fact, a recent consensus has been that network analysis should play a critical role in investigating the paradigm. We hope this relatively simple network analysis reported here will induce more extensive applications of network analysis approaches to the study of human milk microbiome, especially the stability (balance) of the milk ecosystem, which should greatly enhance our understanding of the health and disease implications of milk bacterial microbiota.

There remain some important follow-up questions to this network analysis, and the answers will need efforts from the biomedical research community. Indeed, the insights from our network analysis should only be a starting point for thoroughly understanding the health and disease implications of the human milk microbiome. Experimental and clinical studies of human milk microbiome are crucial, but also understandably difficult to conduct. We realize that even the basic research of the species identification of human milk microbes is still at the very preliminary stage. Both our analysis and the study done by Hunt et al (2011)[1], from which we obtained data for reconstructing the milk microbiome network, only reported the genera of bacteria present, but not the species within these genera. This data limitation strongly constrained our capability to infer further information on the bacterial interaction within human milk microbiome. Given the self-evident, critical importance of human milk microbiome, we hope this study will be a stepping stone for more advanced network analysis based on more comprehensive biomedical data sets generated by future studies of the human milk microbiome.

## Acknowledgements

We appreciate Prof. Zhang Yaping, Academician and Vice President of the Chinese Academy of Science (CAS), for reviewing the manuscript and for his insightful comments and suggestions, which played a significant role in our preparing for this submission. Z. Ma's research received funding from the following grants: "CAS One-Hundred Talented PI Program"（中科院百人计划）, "Exceptional Scientists Program of Yunnan Province"（云南省高端科技人才）and National Science Foundation of China (NSFC Grant No: 61175071).